\documentclass[a4paper,11pt]{article}
\usepackage{pos}

\def\beq{\begin{equation}}
\def\eeq{\end{equation}}
\def\beqa{\begin{eqnarray}}
\def\eeqa{\end{eqnarray}}

\title{Theoretical calculations for $t{\bar t}H$ production}

\author*[a]{Nikolaos Kidonakis}
\author[b]{Nodoka Yamanaka}

\affiliation[a]{Department of Physics, Kennesaw State University,\\
Kennesaw, GA 30144, USA}
\affiliation[b]{Nishina Center for Accelerator-Based Science, RIKEN,\\
Wako 351-0198, Japan}

\emailAdd{nkidonak@kennesaw.edu}
\emailAdd{nodoka.yamanaka@riken.jp}

\abstract{We calculate higher-order perturbative corrections for $t{\bar t}H$ production. By adding second-order and third-order soft-gluon corrections to NLO QCD and NLO electroweak (EW) results, we present total and differential cross sections at LHC energies through approximate N$^3$LO (aN$^3$LO) in QCD and NLO in EW. Our aN$^3$LO QCD + NLO EW results provide significant enhancements and a reduced scale dependence for the cross section and the top-quark transverse-momentum and rapidity distributions.}

\FullConference{43rd International Conference on High Energy Physics (ICHEP 2026)\\
30 July to 5 August , 2026\\
Natal, Brazil\\}

\begin{document}
\maketitle

\section{Introduction}

The production of a top-antitop quark pair and a Higgs boson ($t{\bar t}H$) involves the two heaviest particles in the Standard Model, and it enables measurements of the Yukawa coupling of the top quark. The process has been observed by the ATLAS \cite{ATLAS} and CMS \cite{CMS} collaborations at the LHC, and cross section measurements have been made at 13 TeV energy.

Theoretically, there are significant contributions from higher-order perturbative corrections, and many calculations have been performed over the years for both the strong and the weak interactions (see \cite{NKNY} for a list of references). To achieve better precision, calculations are needed beyond NLO. Here, we derive results that build on NLO QCD \cite{NLO} calculations by adding second-order and third-order soft-gluon corrections, and we present cross sections and differential distributions at aN$^3$LO QCD + NLO EW \cite{NKNY}.

In Section 2, we present briefly the theoretical formalism of soft-gluon resummation for the derivation of higher-order soft-gluon corrections \cite{NKGS,NKtop}. In Section 3, we present numerical results for the total cross section at 13, 13.6, and 14 TeV LHC energies. In Section 4, we display the top quark transverse momentum, $p_T$, and rapidity distributions. In Section 5, we derive a constraint on the top quark Yukawa coupling. We conclude in Section 6.

\section{Higher-order soft-gluon corrections}

Soft-gluon corrections are important for processes that involve the top quark, in particular the production of top-antitop pairs alone or in association with electroweak or Higgs bosons \cite{NKNY,NKtop,NKttV}, since the large mass brings the process closer to partonic threshold. These corrections are numerically dominant and they approximate known exact results at NLO and NNLO very well. 

The lowest-order partonic processes for $t{\bar t}H$ production are $ a(p_a)\, + \, b\, (p_b) \to t(p_1)\, + \, {\bar t}(p_2) \, + \, H(p_3)  $ with $a$ and $b$ denoting gluons or quarks and antiquarks. We use the kinematical variables $s=(p_a+p_b)^2$, $s'=(p_1+p_2)^2$, $t=(p_a-p_1)^2$, $t'=(p_b-p_2)^2$, $u=(p_b-p_1)^2$, $u'=(p_a-p_2)^2$, and we define the partonic threshold variable $s_4=s+t+u-m_t^2-m_{{\bar t} H}^2$. At partonic threshold $s_4 \to 0$.

The soft-gluon corrections appear in the perturbative series as plus distributions of logarithms involving $s_4$, i.e. $[(\ln^k(s_4/m_t^2))/s_4]_+$ with the power $k \le 2n-1$ for the order $\alpha_s^n$ corrections. We take Laplace transforms and resum these soft corrections for the double-differential cross section in one-particle-inclusive (1PI) kinematics. 

The cross section in integral tranform space can be factorized into functions that describe universal soft and collinear emission from the incoming partons, a hard function $H_{ab \to t{\bar t}H}$ that describes the hard scattering, and a soft function $S_{ab \to t{\bar t}H}$ for noncollinear soft-gluon emission. The renormalization group evolution of the parton functions and the soft function results in the resummation of soft-gluon corrections in the cross section as a function of the transform variable $N$, and we find
\beqa
{\hat{\sigma}}_{ab \to t{\bar t}H}^{res}(N) &=&   
\exp\left[ \sum_{i=a,b} E_i(N_i)\right] \; 
\exp \left[\sum_{i=a,b} 2\int_{\mu_F}^{\sqrt{s}} \frac{d\mu}{\mu}\;
\gamma_{i/i}\left({\tilde N}_i,\alpha_s(\mu)\right)\right] \;
\nonumber\\ && \hspace{-8mm} \times \,
{\rm tr} \left \{H_{ab \to t{\bar t}H} \left(\alpha_s(\sqrt{s})\right) \;
{\bar P} \, \exp \left[\int_{\sqrt{s}}^{{\sqrt{s}}/{\tilde N}} \frac{d\mu}{\mu} \;
\Gamma^{\dagger}_{S \, ab \to t{\bar t}H} \left(\alpha_s(\mu)\right)\right] \right.
\nonumber\\ && \left. \times \,
S_{ab \to t{\bar t}H} \left(\alpha_s(\sqrt{s}/{\tilde N}) \right) \;
P \, \exp \left[\int_{\sqrt{s}}^{{\sqrt{s}}/{\tilde N}} 
\frac{d\mu}{\mu}\; \Gamma_{S \, ab \to t{\bar t}H}
\left(\alpha_s(\mu)\right)\right] \right\} \, .
\eeqa
The first exponent resums universal terms for collinear and soft emission while the second term provides the factorization scale dependence. 

The soft anomalous dimension matrix, $\Gamma_{S \, q{\bar q} \to t{\bar t}H}$, for the process $q{\bar q} \to t{\bar t}H$ is a $2\times 2$ matrix.
We choose an $s$-channel singlet and octet exchange color basis, i.e. $\delta_{ab} \, \delta_{12}$ and $T^c_{ba} T^c_{12}$.
At one loop, we have the following expressions for the four matrix elements:
\beqa
&& \hspace{-7mm} \Gamma_{11 \, q{\bar q} \to t{\bar t}H}^{(1)}=\Gamma_{\rm cusp}^{(1)},
\quad
\Gamma_{12 \, q{\bar q} \to t{\bar t}H}^{(1)}=
\frac{C_F}{2N_c} \Gamma_{21 \, q{\bar q} \to t{\bar t}H}^{(1)},
\quad
\Gamma_{21 \, q{\bar q} \to t{\bar t}H}^{(1)}=
\ln\left(\frac{(t-m_t^2)(t'-m_t^2)}{(u-m_t^2)(u'-m_t^2)}\right),
\nonumber \\ && \hspace{-7mm}
\Gamma_{22 \, q{\bar q} \to t{\bar t}H}^{(1)}=\left(1-\frac{C_A}{2C_F}\right)
\left[\Gamma_{\rm cusp}^{(1)}+2C_F\ln\left(\frac{(t-m_t^2)(t'-m_t^2)}{(u-m_t^2)(u'-m_t^2)}\right)\right]
\nonumber \\ && \hspace{13mm}
{}+\frac{C_A}{2}\left[\ln\left(\frac{(t-m_t^2)(t'-m_t^2)}{s\, m_t^2}\right)-1\right]
\eeqa
where the cusp anomalous dimension is a function of $\beta'=\sqrt{1-4m_t^2/s'}$.

At two loops, the four matrix elements are 
\beqa
&& \Gamma_{11 \, q{\bar q} \to t{\bar t}H}^{(2)}=\Gamma_{\rm cusp}^{(2)}, \quad \quad
\Gamma_{12 \, q{\bar q} \to t{\bar t}H}^{(2)}=
\left(K_2-C_A \, N_2\right) \Gamma_{12 \, q{\bar q} \to t{\bar t}H}^{(1)}, 
\nonumber \\ && 
\Gamma_{21 \, q{\bar q} \to t{\bar t}H}^{(2)}=
\left(K_2+C_A \, N_2\right) \Gamma_{21 \, q{\bar q} \to t{\bar t}H}^{(1)},
\nonumber \\ && 
\Gamma_{22 \, q{\bar q} \to t{\bar t}H}^{(2)}=
K_2 \, \Gamma_{22 \, q{\bar q} \to t{\bar t}H}^{(1)}
+\left(1-\frac{C_A}{2C_F}\right)
\left(\Gamma_{\rm cusp}^{(2)}-K_2 \, \Gamma_{\rm cusp}^{(1)}\right)
+\frac{1}{4} C_A^2(1-\zeta_3). 
\eeqa
Expressions for $\Gamma_{\rm cusp}^{(1)}$, $\Gamma_{\rm cusp}^{(2)}$, $K_2$, and $N_2$ can be found in Refs. \cite{NKGS,NKtop,NKttV}.

The soft anomalous dimension matrix, $\Gamma_{S \, gg \to t{\bar t}H}$, for the process $gg\to t{\bar t}H$ is a $3\times 3$ matrix. We choose the color basis $\delta_{ab} \, \delta_{12}$, $d^{abc} T^c_{12}$, $if^{abc} T^c_{12}$. At one loop, we have the following expressions for the nine matrix elements:
\beqa
&& \hspace{-5mm} \Gamma_{11 \, gg \to t{\bar t}H}^{(1)}= \Gamma_{\rm cusp}^{(1)}, \quad \quad \Gamma_{12 \, gg \to t{\bar t}H}^{(1)}=\Gamma_{21 \, gg \to t{\bar t}H}^{(1)}=0, 
\nonumber \\ && \hspace{-5mm} 
\Gamma_{13 \, gg \to t{\bar t}H}^{(1)}= \frac{1}{2}\ln\left(\frac{(t-m_t^2)(t'-m_t^2)}{(u-m_t^2)(u'-m_t^2)}\right), \quad \quad \Gamma_{31 \, gg \to t{\bar t}H}^{(1)} = 2 \, \Gamma_{13 \, gg \to t{\bar t}H}^{(1)},
\nonumber \\ && \hspace{-5mm} 
\Gamma_{22 \, gg \to t{\bar t}H}^{(1)}= \left(1-\frac{C_A}{2C_F}\right) \Gamma_{\rm cusp}^{(1)}
+\frac{C_A}{2}\left[\frac{1}{2}\ln\left(\frac{(t-m_t^2)(t'-m_t^2)(u-m_t^2)(u'-m_t^2)}{s^2\, m_t^4}\right)-1\right],
\nonumber \\ && \hspace{-5mm}
\Gamma_{23 \, gg \to t{\bar t}H}^{(1)}=\frac{C_A}{2} \, \Gamma_{13 \, gg \to t{\bar t}H}^{(1)}, \quad \quad
\Gamma_{32 \, gg \to t{\bar t}H}^{(1)}=\frac{(N_c^2-4)}{2N_c} \, \Gamma_{13 \, gg \to t{\bar t}H}^{(1)}, 
\nonumber \\ && \hspace{-5mm}
\Gamma_{33 \, gg \to t{\bar t}H}^{(1)}=\Gamma_{22 \, gg \to t{\bar t}H}^{(1)}.
\eeqa
At two loops, the nine matrix elements are 
\beqa
&& \hspace{-8mm} \Gamma_{11 \, gg \to t{\bar t}H}^{(2)}= \Gamma_{\rm cusp}^{(2)}, \quad \quad \Gamma_{12 \, gg \to t{\bar t}H}^{(2)}=\Gamma_{21 \, gg \to t{\bar t}H}^{(2)}=0, 
\nonumber \\ && \hspace{-8mm}
\Gamma_{13 \, gg \to t{\bar t}H}^{(2)}=\left(K_2-C_A N_2\right) \Gamma_{13 \, gg \to t{\bar t}H}^{(1)}, \quad \quad 
\Gamma_{31 \, gg \to t{\bar t}H}^{(2)}=\left(K_2+C_A N_2\right) \Gamma_{31 \, gg \to t{\bar t}H}^{(1)},
\nonumber \\ && \hspace{-8mm}
\Gamma_{22 \, gg \to t{\bar t}H}^{(2)}= K_2 \, \Gamma_{22 \, gg \to t{\bar t}H}^{(1)}
+\left(1-\frac{C_A}{2C_F}\right) \left(\Gamma_{\rm cusp}^{(2)}-K_2 \Gamma_{\rm cusp}^{(1)}\right)+\frac{C_A^2}{4}(1-\zeta_3),
\nonumber \\ && \hspace{-8mm}
\Gamma_{23 \, gg \to t{\bar t}H}^{(2)}= K_2 \, \Gamma_{23 \, gg \to t{\bar t}H}^{(1)}, \quad
\Gamma_{32 \, gg \to t{\bar t}H}^{(2)}= K_2 \, \Gamma_{32 \, gg \to t{\bar t}H}^{(1)}, \quad
\Gamma_{33 \, gg \to t{\bar t}H}^{(2)}=\Gamma_{22 \, gg \to t{\bar t}H}^{(2)}.  
\eeqa

We perform finite-order expansions of the resummed cross section and invert back to momentum space with no need for a prescription. By adding the second-order and third-order soft-gluon corrections to exact NLO results, we derive, respectively, approximate NNLO (aNNLO) and approximate N$^3$LO (aN$^3$LO) predictions for cross sections and top-quark differential distributions in transverse momentum and rapidity.

\section{$t{\bar t}H$ cross sections}

\begin{table}[htbp]
\begin{center}
\begin{tabular}{|c|c|c|c|c|c|c|} \hline
\multicolumn{4}{|c|}{$t{\bar t}H$ cross sections in $pp$ collisions at the LHC} \\ \hline
$\sigma$ in fb & 13 TeV & 13.6 TeV & 14 TeV \\ \hline
LO QCD & $434^{+138}_{-97}{}^{+9}_{-9}$ & $485^{+152}_{-108}{}^{+10}_{-10}$ & $520^{+162}_{-115}{}^{+10}_{-11}$ \\ \hline
NLO QCD & $497^{+14}_{-38}{}^{+11}_{-12}$ & $558^{+17}_{-43}{}^{+12}_{-13}$ & $601^{+18}_{-46}{}^{+13}_{-14}$ \\ \hline
NLO QCD + NLO EW & $503^{+15}_{-37}{}^{+11}_{-12}$ & $564^{+17}_{-42}{}^{+13}_{-13}$ & $607^{+19}_{-45}{}^{+13}_{-14}$ \\ \hline
aNNLO QCD & $518^{+4}_{-14}{}^{+12}_{-13}$ & $581^{+4}_{-15}{}^{+13}_{-14}$ & $626^{+5}_{-16}{}^{+14}_{-15}$ \\ \hline
aNNLO QCD + NLO EW & $524^{+5}_{-13}{}^{+12}_{-13}$ & $587^{+5}_{-14}{}^{+13}_{-14}$ & $632^{+6}_{-15}{}^{+14}_{-15}$ \\ \hline
aN$^3$LO QCD & $523^{+3}_{-7}{}^{+12}_{-13}$ & $587^{+3}_{-7}{}^{+13}_{-14}$ & $633^{+3}_{-8}{}^{+14}_{-15}$ \\ \hline
aN$^3$LO QCD + NLO EW & $529^{+3}_{-6}{}^{+12}_{-13}$ & $593^{+3}_{-6}{}^{+13}_{-14}$ & $639^{+3}_{-7}{}^{+14}_{-15}$ \\ \hline
\end{tabular}
\caption[]{The $t{\bar t}H$ cross sections (in fb) in $pp$ collisions at LHC energies with QCD and EW corrections at different perturbative orders, with scale and pdf uncertainties, with $m_t=172.5$ GeV, $m_H=125.2$ GeV, and MSHT20 aN$^3$LO pdf.}
\label{table1}
\end{center}
\end{table}

In Table 1, we present total cross sections for $t{\bar t}H$ production in proton-proton  collisions at LHC energies of 13, 13.6, and 14 TeV. The central results are with the factorization and renormalization scales set equal to the top quark mass, $\mu=m_t=172.5$ GeV, with a Higgs mass $m_H=125.2$ GeV, and with MSHT20 aN$^3$LO \cite{MSHT20} parton distribution functions (pdf). The first uncertainty is from scale variation by a factor of 2, and the second uncertainty is from the pdf.

At 13.6 TeV, the NLO QCD corrections increase the LO result by 15\%; the aNNLO QCD corrections provide an additional 5\%, and the aN$^3$LO QCD corrections add a further 1\%. We observe a big reduction in scale uncertainty at higher orders.
Similar observations are made for the 13 TeV and 14 TeV cross sections.

\section{Top quark $p_T$ and rapidity distributions in $t{\bar t}H$ production}

We continue our study of $t{\bar t}H$ production by calculating top quark differential distributions for this process. We present results at perturbative orders from LO QCD through aN$^3$LO QCD + NLO EW.

\begin{figure}[htbp]
\begin{center}
\includegraphics[width=68mm]{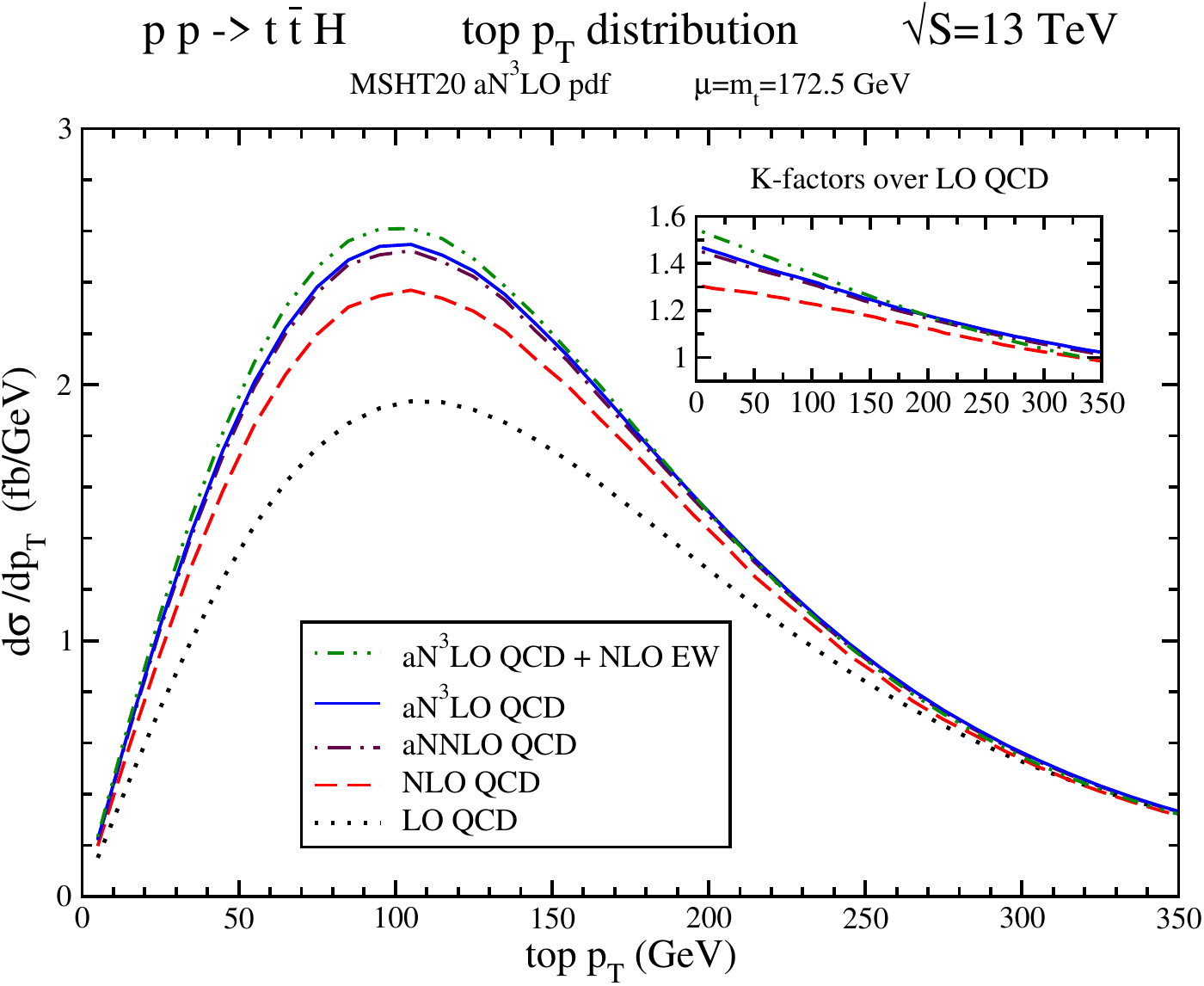}
\hspace{5mm}
\includegraphics[width=68mm]{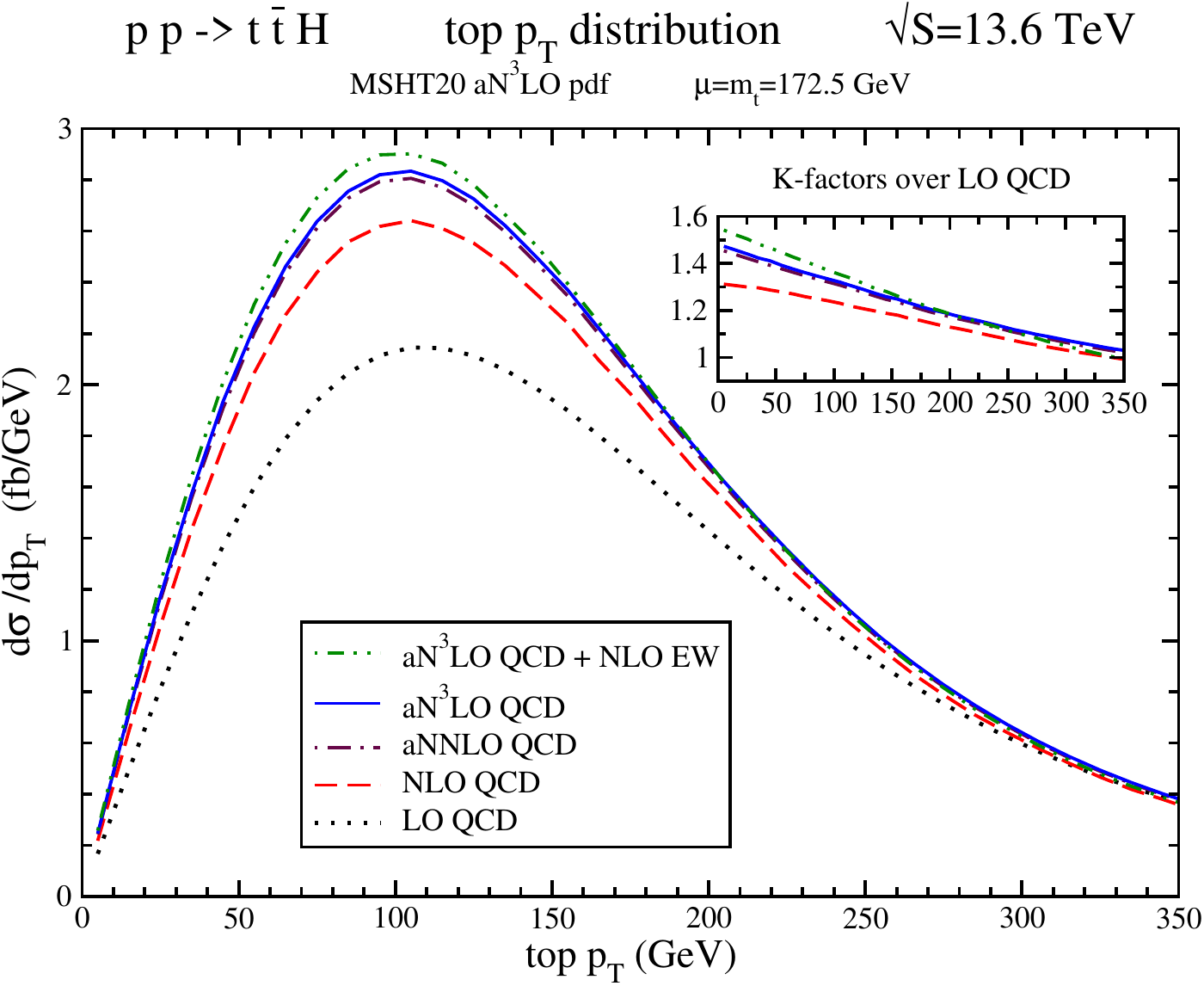}
\caption{Top quark $p_T$ distributions at 13 TeV (left) and 13.6 TeV (right).}
\label{toppT}
\end{center}
\end{figure}

In Fig. \ref{toppT}, we display the top quark $p_T$ distributions at 13 and 13.6 TeV collider energies. We observe that the $K$-factors, which are shown in the inset plots, are large at small $p_T$ and gradually decrease at large $p_T$.

\begin{figure}[htbp]
\begin{center}
\includegraphics[width=68mm]{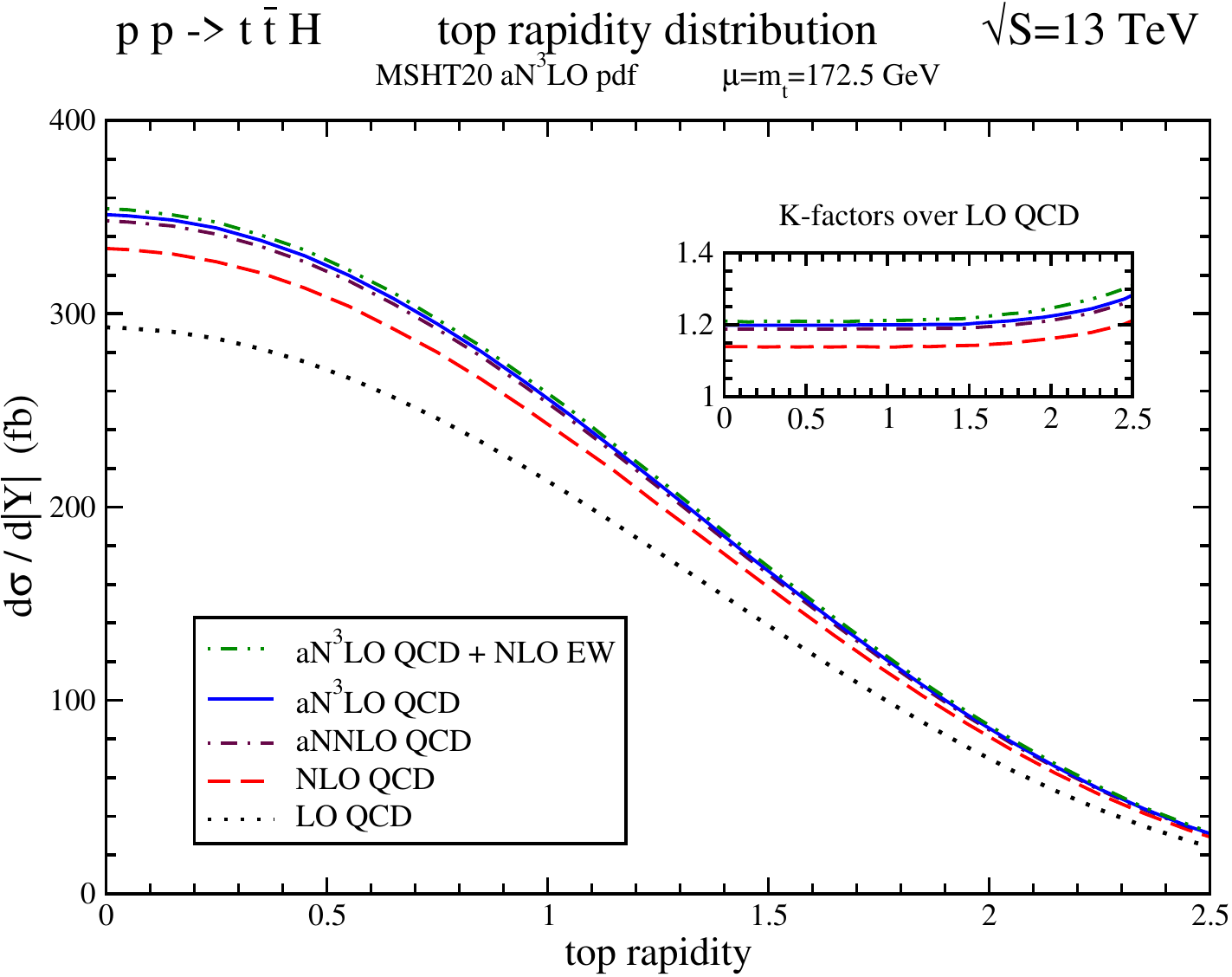}
\hspace{5mm}
\includegraphics[width=68mm]{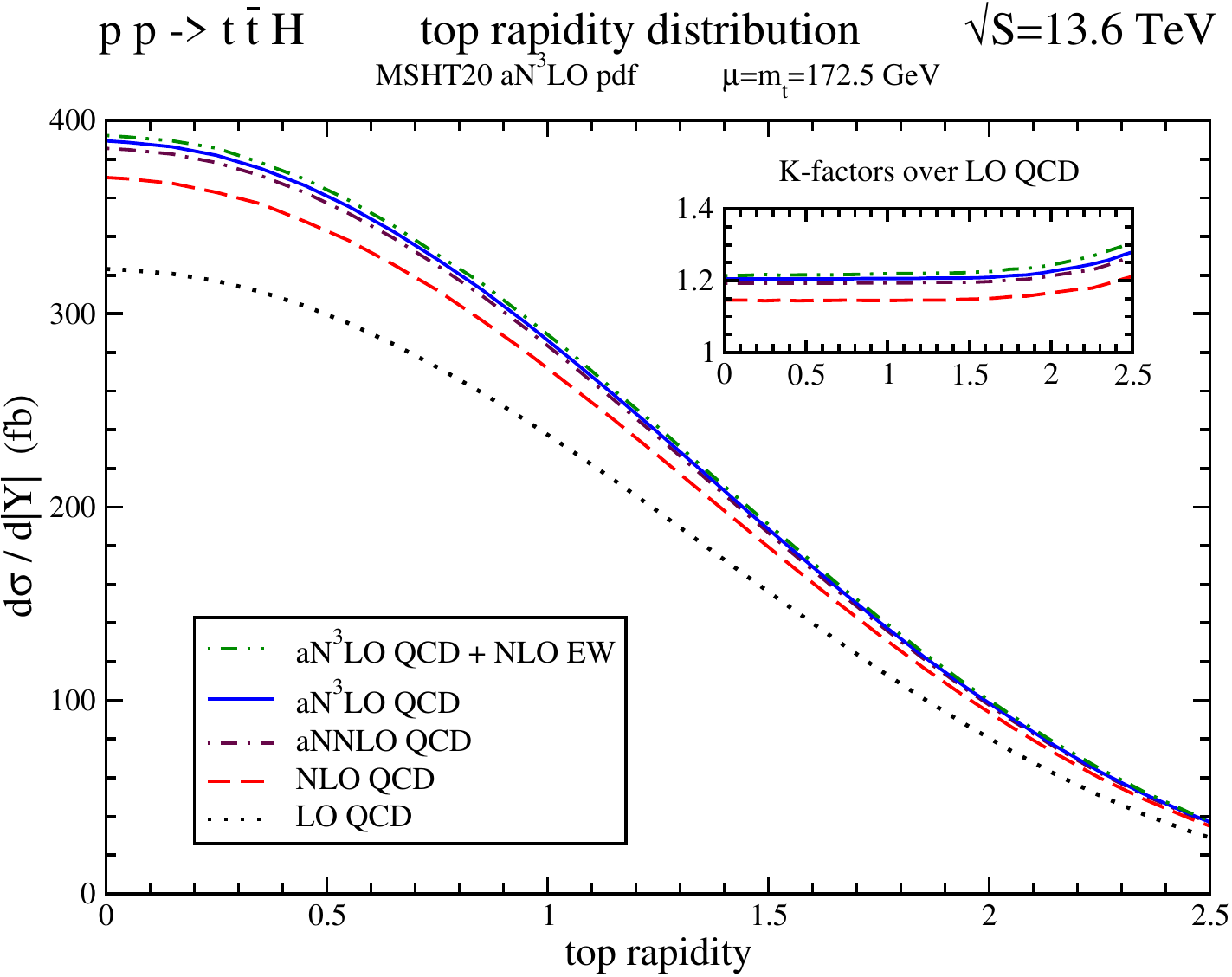}
\caption{Top quark rapidity distributions at 13 TeV (left) and 13.6 TeV (right).}
\label{topy}
\end{center}
\end{figure}

In Fig. \ref{topy}, we display the top quark rapidity distributions at 13 and 13.6 TeV collider energies. We observe that the $K$-factors increase significantly at larger rapidities.

\section{$t{\bar t}H$ production and top-quark Yukawa coupling}

The $t{\bar t}H$ cross section is proportional to the square of the top-quark Yukawa coupling, $Y_t^ 2$. Using our results for the cross section, we can derive an upper limit to its deviation $\delta Y_t$ from the Standard Model value. A comparison with the latest ATLAS data in \cite{ATLAS} gives $|\delta Y_t| \, \le \, 0.44 \, Y_t$.

\section{Summary}

We have calculated higher-order corrections for $t{\bar t}H$ production. We have used soft-gluon resummation and N$^3$LO expansions to provide higher-order results. The soft-gluon corrections are dominant and we make aN$^3$LO QCD + NLO EW calculations for cross sections and differential distributions. These higher-order corrections further enhance and improve the theoretical predictions. Finally, we set an upper limit on the deviation of the top quark Yukawa coupling from its Standard Model value.

\acknowledgments

The work of N.K. is supported by the National Science Foundation under Grant No. PHY 2412071. The work of N.Y. is supported by the RIKEN TRIP initiative (Nuclear transmutation), and also partially by ERATO project-JPMJER2304.

\end{document}